\documentclass[pdflatex,sn-mathphys-num]{sn-jnl}

\usepackage{graphicx}
\usepackage{subcaption}
\usepackage{graphicx}%
\usepackage{multirow}%
\usepackage{amsmath,amssymb,amsfonts}%
\usepackage{amsthm}%
\usepackage{mathrsfs}%
\usepackage[title]{appendix}%
\usepackage{xcolor}%
\usepackage{textcomp}%
\usepackage{manyfoot}%
\usepackage{booktabs}%
\usepackage{algorithm}%
\usepackage{algorithmicx}%
\usepackage{algpseudocode}%
\usepackage{listings}%

\theoremstyle{thmstyleone}%
\theoremstyle{thmstyletwo}%

\theoremstyle{thmstylethree}%

\begin{document}

\title[Article Title]{A tapestry of gravitational waveforms to reveal black hole properties}


\author*[1]{\fnm{Zack} \sur{Carson}}\email{zackrcarson@gmail.com}

\affil*[1]{\orgname{United States Space Force}, \orgaddress{\city{Colorado Springs}, \state{CO}, \country{USA}}}

\abstract{A novel approach to binary black hole gravitational wave analysis improves the process of inferring black hole properties by selecting the most accurate waveform model for each region of the parameter space, resulting in tighter constraints and 30\% lower computational costs.}

\maketitle

\begin{figure}[htbp]


  \centering
  \includegraphics[width=\textwidth]{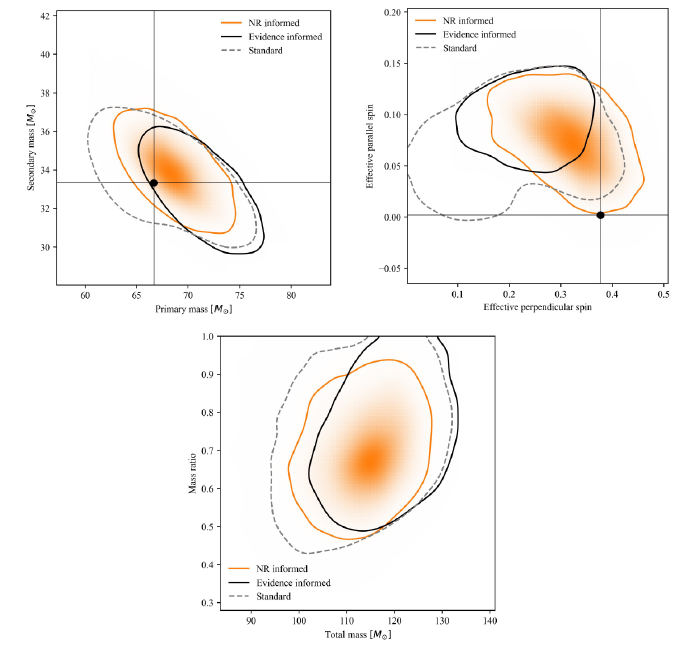}

  \caption{Two-dimensional posterior probability distributions obtained in the analysis of the SXS:BBH:0926 and GW191109\_010717 events. The top panels show the measured individual BH masses and spin components in the simulated event, and the bottom panel shows the measurement of the total binary mass vs. the mass ratio for the real event. All contours represent 90\% credible intervals, and the black cross hairs indicate the true value.}
  \label{fig:posteriors}
\end{figure}

Inferring the properties of distant binary black hole (BH) systems via the gravitational radiation released during violent inspiral and merger events is a crucial study to our understanding of the universe. Accurately interpreting these events taking place in the most extreme regions of spacetime in existence--where gravity is exceptionally strong, non-linear, and highly dynamical--gives physicists key insights into the nature of fundamental physics, in a new laboratory not accessible anywhere else in the universe. In such complex environments, care must be taken when modeling the resulting gravitational waves (GWs) following Einstein's general relativity (GR), else systematic uncertainties arising from the model's mismatch with reality will dominate the overall error. The work done by Hoy et al.~\cite{Hoy2025} introduces a new method to reduce these uncertainties by prioritizing the most accurate models in each region of the parameter space, based on their fidelity to numerical relativity (NR) simulations, computational approximations of true solutions to Einstein's equations. Applying the new ``\textit{NR informed}'' method to both simulated and real GW data, the authors demonstrate this new method to more faithfully capture the true BH mass and spin than the standard method currently utilized by the LIGO-Virgo-KAGRA (LVK) collaboration~\cite{Abbott2021}, while also reducing computational cost by 30\%.

Measuring the properties, such as mass and spin, of binary BH mergers is an extremely difficult task due to the complexity of solving Einstein's equations - something that takes millions of CPU hours on high-performance compute clusters~\cite{Hamilton2024} to obtain NR approximations. From the handful of models available to use, each of which is known to be more faithful to general relativity in different regions of parameter space~\cite{MacUilliam2024Survey}, scientists historically perform Bayesian analyses across the full waveform with each selected model individually, finally combining the results at the end using a variety of weighting mechanisms. Hoy et al. improves upon this process by instead performing a single multi-model Bayesian analysis. By estimating the accuracy of several GW models across the parameter space using mismatch comparisons with high-fidelity NR simulations,  parameter-dependent weights can be assigned to each model, favoring the most accurate in each parameter space region. This allows one to combine models into a single Bayesian inference, reducing bias and computational cost while injecting model-specific accuracies into the analysis for the first time. Another key feature of this ``\textit{NR informed}'' method is its flexibility: it can utilize any combination of GW models as more accurate ones become available.

To demonstrate the success of the new approach, Hoy et al. first applied it to the simulated BH merger SXS:BBH:0926~\cite{Boyle2019, Blackman2017}, combining three of the most accurate, cutting-edge GW models currently in use~\cite{Pratten2021, Estelles2022, RamosBuades2023} as discussed above. The top row of Fig.~\ref{fig:posteriors} shows the resulting two-dimensional posterior distributions on the BH masses and spins using the \textit{NR informed} method presented by Hoy et al., as well as the \textit{Evidence Informed} and \textit{Standard} methods which combine separate analyses with different GW models--the latter currently being used by the LVK collaboration. Observing the mass distribution, all three methods contain the true values at the 90\% credible interval, while both the \textit{NR informed} and \textit{Standard} methods also do so at the 50\% level. In the spin distribution, significant differences in the posteriors can be seen, with only Hoy et al.'s \textit{NR informed} method correctly capturing the true values at a 90\% credible interval. Further, due to the single multi-model Bayesian inference used by Hoy et al., the full analysis only took a total of 230 CPU days, compared to the 334 CPU days required by existing methods requiring each inference to be run separately.

Hoy et al. also applied their \textit{NR informed} method to the real GW signal GW191109\_010717~\cite{Abbott2021}, observed on November 9, 2019 with an estimated total mass  of $92 M_{\odot} < M < 125 M_{\odot}$. This event generated significant interest because it likely involved large BH masses from the ``upper mass gap'', theorized to have been formed in hierarchical merger processes to create BHs with masses larger than $65M_{\odot}$, the predicted limit for stellar collapse BHs~\cite{Woosley2002}. The authors were able to more tightly constrain the total mass to $100 M_{\odot} < M < 124 M_{\odot}$, and provided strong evidence for unequal source masses as shown in the bottom row of Fig.~\ref{fig:posteriors}. This re-analysis increased the probability that the primary component mass lies within the upper mass gap from 51\% in the original LVK analysis~\cite{Abbott2021} to 69\%--further supporting the hypothesis that the primary BH was indeed formed through a hierarchical merger process.

Hoy et al.'s work presents a major advancement in the analysis of GW signals from binary BH mergers, demonstrating a new technique that incorporates model uncertainty into Bayesian inference for the first time; both outperforming traditional methods and reducing computational cost. This important work heralds in a new era of GW analysis where scientists can utilize different GW models in a modular, ``plug and play'', fashion. While there is no guarantee that each chosen model is accurate enough to avoid biases in the final results, this method is still the most accurate of those currently used, and continuous improvements can always be made by swapping out individual models with increasingly more accurate ones--which are always being developed independently by GW researchers around the world. The potential of Hoy et al.'s new approach will only grow as the field matures and better models are developed, positioning it to become an essential framework in the future of GW astronomy, allowing for more reliable insights into the universe’s most extreme objects and events.

\bibliography{sn-article}

\section*{Competing interests}
The author declares no competing interests.

\end{document}